\documentclass[letterpaper,twocolumn,10pt]{article}
\usepackage{usenix,epsfig,endnotes}
\usepackage{hyperref}
\usepackage{url}

\begin{document}

%don't want date printed
\date{}

%make title bold and 14 pt font (Latex default is non-bold, 16 pt)
\title{\Large \bf How Do Organizations Seek Cyber Assurance? 

Investigations on the Adoption of the Common Criteria and Beyond}
\author{
{\rm Nan Sun}\\
Deakin University
\and
{\rm Chang-Tsun Li}\\
Deakin University
\and
{\rm Hin Chan}\\
Australian Cyber Security Centre
\and
{\rm Md Zahidul Islam}\\
Charles Sturt University
\and
{\rm Md Rafiqul Islam}\\
Charles Sturt University
\and
{\rm Warren Armstrong}\\
QuintessenceLabs Pty Ltd.
}

%\author{
%{\rm Nan Sun}\\
%Deakin University
%\and
%{\rm Chang-Tsun Li}\\
%Deakin University
%\and
%{\rm MD Zahidul Islam}\\
%Charles Sturt University
%\and
%{\rm MD Rafiqul Islam}\\
%Charles Sturt University
%\and
%{\rm Hin Chan}\\
%Australia Cyber Security Centre
%\and
%{\rm Warren Armstrong}\\
%QuintessenceLabs Pty Ltd
%}

\maketitle

% Use the following at camera-ready time to suppress page numbers.
% Comment it out when you first submit the paper for review.
\thispagestyle{empty}

\subsection*{Abstract}

Cyber assurance, which is the ability to operate under the onslaught of cyber attacks and other unexpected events, is essential for organizations facing inundating security threats on a daily basis. Organizations usually employ multiple strategies to conduct risk management to achieve cyber assurance. Utilizing cybersecurity standards and certifications can provide guidance for vendors to design and manufacture secure Information and Communication Technology (ICT) products as well as provide a level of assurance of the security functionality of the products for consumers. Hence, employing security standards and certifications is an effective strategy for risk management and cyber assurance. In this work, we begin with investigating the adoption of cybersecurity standards and certifications by surveying 258 participants from organizations across various countries and sectors. Specifically, we identify adoption barriers of the Common Criteria through the designed questionnaire. Taking into account the seven identified adoption barriers, we show the recommendations for promoting cybersecurity standards and certifications. Moreover, beyond cybersecurity standards and certifications, we shed light on other risk management strategies devised by our participants, which provides directions on cybersecurity approaches for enhancing cyber assurance in organizations.  

%Furthermore, we also shed lighton other risk management strategies devised by our par-ticipants beyond the Common Criteria and cybersecuritystandards as well as certifications, which underscores theimportance  of  interweaving  multi-layered  complemen-tary cybersecurity approaches for enhancing cyber assur-ance.
\section{Introduction}
According to the statistics from Dell Technology in 2019/2020, 44\% of organizations have experienced at least one cybersecurity attack or data breach during the prior twelve months \cite{dellt}. Security issues are becoming a daily struggle for public and private sectors alike. Data from the Australian Cyber Security Centre's (ACSC) Annual Cyber Threat Report \cite{ACSA} shows that the number of cybersecurity attacks is still on the rise. As the consequences of a cybersecurity attack, an organization's financial and reputational health may be affected, business operations are disrupted, sensitive information including intellectual property may be stolen, and malicious activity may continue \cite{sun2018data}.

Although it is difficult to quantify the costs of impacts, cybersecurity remediation can be more expensive than early and ongoing investment in prevention \cite{ACSA2}. To reduce the potential impact of cyber attacks, risk management that involves the process of identifying, assessing, and taking steps to minimize security risks is essential as a cybersecurity approach for organizations. Having cybersecurity awareness and robust security strategies in place can help organizations prepare for, protect against, and respond to cyber attacks to some extent \cite{matheu2020survey}. Since Information and Communication Technology (ICT) products are widely used by organizations and individual users, choosing trusted ICT products is of paramount importance for the organization's risk management. The existing cybersecurity standards and certifications for evaluating ICT products provide guidelines for vendors to design, develop, evaluate, certify their products, as well as provide trusted references for users to choose the products.

The Common Criteria for Information Technology Security Evaluation (often referred to as Common Criteria or CC) is an international standard (ISO/IEC 15408), which certifies that systems and products to ensure they meet predefined security requirements \cite{herrmann2002using}. The Common Criteria covers comprehensive ICT security-related technologies and a wider range of evaluation aspects regarding security functionalities and security assurance. Security requirements for a class of related products are typically predefined in a Protection Profile by a user group or user \cite{herrmann2002using}. The purpose of a Protection Profile is to provide reusable templates of security requirements to support the definition of functional security standards and guide the formulation of product development and procurement specifications. 

Generally, the subject of the evaluation that can be part of the product or system is called the Target of Evaluation. %It is possible to use more than one protection profile at a time, depending on the target technology to be certified. 
A Security Target is a document that identifies the security features of the Target of Evaluation \cite{CC}. If a vendor has an ICT product that they would like to be evaluated and certified under the Common Criteria, they must complete a Security Target description. The vendor should conduct a self-assessment on compliance with the Protection Profile prior to evaluation against the profile. Evaluation are conducted in laboratories to validate the product's security features and confirm that it meets the security requirements outlined in the Security Target \cite{laboratories}. Following the evaluation of ICT products and systems via a set of specifications and guidelines, the products that passed the evaluation are awarded the Common Criteria certification \cite{stallings2012computer} and be listed on the Common Criteria portal \cite{CC_portal}.

The Common Criteria certification assures consumers that the products they invest in provide reliable security protection for their operational environment and conform to the vendor's claims. Furthermore, the Common Criteria certification increases the competitiveness of vendors' products when consumers compare them to similar products on the market. For government agencies, the Common Criteria certification not only facilitates procurement but increases the transparency of ICT products' security features, facilitating market supervision and surveillance. However, there is a lack of widespread adoption of evaluated ICT products with security functionality by organizations, including governments and commercial sectors \cite{hearn2004does}. For instance, although Australia is a signatory to the Common Criteria Recognition Arrangement (CCRA), the number of certifications through the Australian Information Security Evaluation Program is trivial when compared to the number of certificated products on Common Criteria's Certified Products List \cite{CC_certifiedProductstatistics}.

In this paper, we aim to identify the adoption barriers for security standards and certifications, especially the one which covers the most extensive category of ICT products, the Common Criteria. Through 258 responses to an online questionnaire from participants from Australian and international organizations, we analyze the organizations' attitudes towards being measured against cybersecurity standards and their adoption of the cybersecurity standards. Our participants also describe risk management strategies, such as reactive and proactive cybersecurity countermeasures and multi-layered risk management approaches, adopted by their organizations to pursue cyber assurance. To achieve our aim, in this paper, we address the following research questions:

\textbf{RQ1:} \textit{What are the adoption barriers of the Common Criteria? }

\textbf{RQ2:} \textit{How to promote the adoption of the Common Criteria?}

\textbf{RQ3:} \textit{How do organizations seek cyber assurance beyond adopting security standards and certifications?}

To answer RQ1, based on the identified adoption barriers from the literature review, we design the questionnaire to investigate the adoption barriers for the Common Criteria in Section 4.1. Recommendations for promoting Common Criteria adoption as well as the other security standards and certifications are presented in Section 4.2towards answering RQ2.  Beyond the security standards and certifications, the adopted approaches to cyber assurance are discussed in Section 5 to address RQ3.

\section{Related Work}

\textbf{Cybersecurity standards and certifications.} The applications and adoption of cybersecurity standards and certifications for ICT products and systems have been explored and discussed in previous studies. As ICT products are designed, developed, and implemented, cybersecurity standards and certifications play a significant role, especially in areas such as the Internet of Things (IoT) \cite{matheu2020survey}, smart grids \cite{leszczyna2018cybersecurity}, and software \cite{kara2012review}. The recent study \cite{matheu2020survey} reviewed cybersecurity standards and certifications for the IoT ecosystem by analyzing the various standards and certifications schemes and the challenges associated with implementing them. Additionally, previous works \cite{bures2018internet, dias2018brief,kuzminykh2018analysis} reviewed the key building blocks for the certification process in the context of security testing and risk assessment in IoT. Along with security certifications for IoT, Leszczyna et al. \cite{leszczyna2018cybersecurity} conducted a study that examined smart grid cybersecurity standards and provided insights into the adoption of cybersecurity standards. Specifically, the work \cite{leszczyna2018cybersecurity} examined 36 cybersecurity-related and 12 privacy-related standards and their adoptions in the area of smart grid. Additionally, Kara et al. \cite{kara2012review} reviewed the Common Criteria in a specific field, which shed light on the Common Criteria's application in secure software development.

Since Common Criteria covers a comprehensive range of categories and technologies for ICT products and services, it promotes the mutual recognition of secure ICT products among a broad range of security standards and certifications \cite{fatima2021survey}. For example, Matheu et al. \cite{matheu2020survey} stated that Common Criteria is the most widely used cybersecurity certification in the IoT field. Furthermore, albeit the fact that Russia is neither a Certificate Authorizing Participant nor a Certificate Consuming Participant of the Common Criteria, the history, structure, and features of the Common Criteria used in the Russian scheme are reviewed in \cite{barabanov2015modern} and \cite{barabanov2014russian}. The Common Criteria standard has also been adopted worldwide by other non-Common Criteria Recognition Arrangement nations, such as China, which has its own certification scheme with their adaption of the Common Criteria standard called GB/T 18336 \cite{hu2019summary}. In spite of the significant role of the Common Criteria in ensuring cybersecurity through security standards and certifications, widespread adoption of the Common Criteria and certified products is still a long way off \cite{hearn2004does}. In this work, we investigate adoption barriers of security standards in the case of the Common Criteria by adopting the survey approach. Aside from the adoption barriers identified from the literature review being validated by our designed questionnaire, we explore other previously overlooked adoption barriers.      

\textbf{Cyber assurance and risk management.} Security risk management refers to the process of \textit{``identifying,  assessing,  and taking steps to reduce security risks to an acceptable level"} according to the definition from the Australian Cyber Security Centre \cite{securityriskmanagement}. Additionally to cybersecurity standards and certifications, organizations employ a variety of risk management strategies, which include physical controls (e.g., alarm systems, biometrics, etc.), technological controls (e.g., firewalls, encryption, etc.), and behavioral controls (e.g., security training, policies and procedures, etc.). The topic of risk management for cybersecurity assurance in different industry sectors has been extensively researched from the perspective of technology \cite{ choo2021multidisciplinary, ghadge2019managing, hoppe2021cyber, meszaros2017introducing}, theoretical perspective \cite{webb2014situation}, as well as the practice perspective \cite{ahmad2021can}. For example, the study by Ghadge et al. \cite{ghadge2019managing} focused on cyber risk management in supply chain contexts by conducting a systematic literature review. A clear understanding of the cyber risk challenges and mitigation strategies helps supply chain managers make informed decisions. The paper by Ahmad et al. \cite{ahmad2021can} provided an in-depth case study of a financial organization and outlined a process model that can be used to increase situation awareness in organizations. 

Furthermore, numerous studies offer valuable insights into better risk management in organizations \cite{biener2015insurability,kure2021asset,laube2017strategic,tounsi2018survey}. For instance, Hoppe et al. \cite{hoppe2021cyber} gathered market insights from 37 recent industry surveys and structured them based on the steps of the risk management process. Through the study, the researchers \cite{hoppe2021cyber} found that a lack of security experts and a strained market were the main obstacles to implementing cyber risk management for small and medium-sized businesses. From the strategic aspects of risk management, Laube et al. \cite{laube2017strategic} systematically reviewed works on cyber risk information sharing, which is proved to be beneficial in providing more edges to the defenders in their races against cyber attackers. Tounsi and Rais \cite{tounsi2018survey} argued that the defenders are required to collect and understand cyber-threat intelligence to cope with the ever-increasing sophistication of cyber threat intelligence. To investigate the role of cyber insurance in risk management, Biener et al. \cite{biener2015insurability} conducted an empirical analysis on the insurability of cyber risks. After assessing the market potential in light of the increasing number of high-profile cyber incidents, they concluded \cite{biener2015insurability} with a positive note on cyber insurance. 

Nevertheless, organizations in various industry verticals are still vulnerable to cyber risks and continue to suffer from damages, such as financial loss, data breach, and even reputation loss, caused by cyber attacks \cite{sun2018data}. As part of our research, we investigate, through questionnaires, how organizations in different countries and industries currently seek cyber assurance and risk management in addition to cybersecurity standards and certifications. Combined with the practice, strategies and insights from the previous research work, we further provide the risk management best practices for cyber assurance at a high level.

\section{Our Investigations}

To investigate how organizations seek cyber assurance and their adoption of security standards, especially the Common Criteria, we tried to find participants across different sectors, countries and organizations of various sizes. We collected responses to our questionnaire through Qualtrics from 22 Sep 2021 to 23 Dec 2021 to seek answers to the three research questions we raised in the Introduction. Our study received ethics approval from Deakin University and Charles Sturt University Human Research Ethics Office with the Reference Number SEBE-2021-38 and Protocol Number H21353, respectively. See survey at: \url{https://github.com/nansunsun/DACCA_Questionnaire/blob/main/DACCA_Questionnaire.pdf}. This section discusses the questionnaire design, data collection, and analysis of the responses.

\begin{table}
\centering
\caption{The number of each sector of participants' organizations belong to. One participant can choose more than one sector. The statistics are based on 158 participants who disclosed the sectors where their organizations operate in.}
\label{sector}
\begin{tabular}{cc}
\hline
Sector name & Count\\
\hline
Defence industry&39\\
Health and social care &19\\
Food and agriculture &3\\
Energy and utilities &21\\
Resources and Mining &5\\
Information Communication Technology &109\\
Manufacturing &16\\
Transportation&13\\
Environment, water and soil&9\\
Education&19\\
Financial and insurance services &15\\
Real estate and insurance services &3\\
Wholesale and retail trade &6\\
Legal services &2\\
Others &56\\
\hline
\end{tabular}
\end{table}

\subsection{Questionnaire Design and Data Collection}

There are 28 questions in the questionnaire, including open-ended and closed-ended questions in the forms of short answers or multiple-choice questions. We first investigated participants' demographics using Q1 - Q4 (Question 1 to Question 4), including the organization's name, countries where the participant's organization operates, the size of the organization, and the sectors wherein the participant's organization conducts its businesses.

For participants from organizations that produce ICT products that may be implemented in hardware, firmware, or software confirmed in Q5, we further investigated their adoption of and attitudes towards adopting the Common Criteria through Q8 and Q10 - Q21. Besides the Common Criteria, we explored whether these ICT product manufacturers adopt any other security standards in Q7 and Q9. In addition, the categories that are relevant to the products produced by the participants' organizations were surveyed in Q6.

For the participants whose organizations use ICT products confirmed in Q22, we surveyed if there are any security certification standards the participants' organizations are looking for when they select the ICT products used or to be used within their organizations in Q24 and those certification standards they have obtained in Q25. Specifically, if the organizations use ICT products with Common Criteria certification, the Evaluation Assurance Levels for the products were investigated in Q26. The categories that are relevant to the products used by the participants’ organizations were surveyed in Q22. In the absence of ICT products with a security certification standard, by adopting the open-ended questions, we surveyed the ways the participants' manage risks associated with potentially poor implementation of security functionality within the products through Q27 and the ways the participants go about seeking assurances in the security functionality of the products in Q28.

Through various avenues, we distributed the questionnaires and collected the responses from the participants across different countries, organizations, and sectors. To ensure our survey reached a wide range of respondents, we used several strategies to identify potential participants. Firstly, the contact information of participants with a track record of participating in IT security standards was collected from the International Common Criteria Conference website \cite{ICCC}. Secondly, we tried to expand the participants' list by searching users from Common Criteria Users Forum \cite{CCUF}, which is a community based around those using the Common Criteria and ISO/IEC 15408 standards. Thirdly, we retrieved the authors from the representative Common Criteria literature (i.e., research paper) as the potential participants and collected their contact details. Furthermore, multiple ICT vendors and companies found in the Common Criteria portal \cite{CC_portal} were included to gain a higher response rate of questionnaires. Last but not least, with the support of the Australian Cyber Security Centre (ACSA), the questionnaires were distributed to the ACSA partners \cite{partner}, which aims to gain the broader viewpoint from participants in
the wider cyber security community.

Email, Linkedin Message, and blog post \footnote{https://www.quintessencelabs.com/blog/quintessencelabs-joins-research-study-with-deakin-university-and-charles-stuart-university/} were used to reach out to participants interested in cybersecurity standards, which included certification bodies, evaluation laboratories, researchers, policymakers, product developers, sellers, and buyers. Participation in the project is voluntary, and participants are free to stop and skip any question at any time if they do not wish to reveal any specific information. The total number of valid responses was 285, of which 177 answered the entire 28 questions.

\begin{figure}
  \centering
  \includegraphics[width=0.9\linewidth]{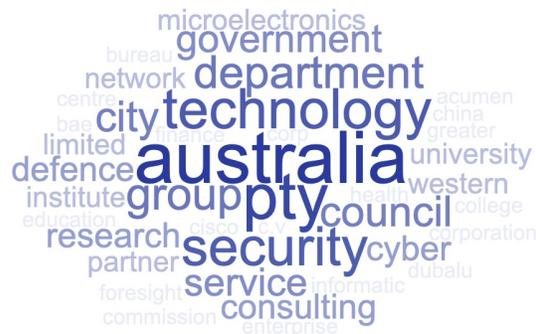}
  \caption{Word cloud of the names of participants' organizations that also infer the operating sectors of organizations. The statistics are based on 152  participants of responses who disclosed the name of their organizations. Words that occur frequently appear larger and darker in colour.}
  \label{wordcloud}
\end{figure}

\subsection{Responses Analysis}

Our participants are from different organizations and sectors, as shown in Table \ref{sector}. 32.54\% of organizations operate in the ICT sector, which is the most significant proportion of participants. Besides the sectors listed in Table \ref{sector}, including the defence industry, energy and utilities, education, etc., 35.44\% of participants chose the ``\textit{Others}" option. Based on the analysis of these participants, it appears that most of them are from city councils, police departments, government departments, and national commissions. A few are from consulting and cybersecurity companies, and the others are anonymous participants. To illustrate the range of sectors of the participants, we present the word cloud of the names of participants' organizations based on the responses for Q1 in Figure \ref{wordcloud}.

Both Australian and international participants are involved in our survey. There are 175 participants' organizations operating in Australia, which takes up 61.40\% of the participants. Second place goes to the United States of America, with 11.93\%. Besides, organizations operating in Asian, North America, Oceania, South American, and African countries are involved in the survey. Note that these statistics do not reflect the popularity of certifying ICT security products in various countries, but to demonstrate that we made a serious attempt to explore beyond Australia and provide a global picture of the certification adoption landscape. 

Lastly, from the perspective of the size of organizations, %we further analyzed the participants of the responses.
the responses are from organizations with various sizes, including large, mid-market, and small businesses. Among the responses, 37.50\% are organizations with more than 1000 employees. Around half (i.e., 46.15\%) of the organizations have 11-1000 employees. Besides, there are 16.35\% organizations with 0-10 employees participating in our survey from the small business.

%Organizations in North America and Oceania, organizations operating in Asian countries (e.g., China, Japan, Singapore, etc.), European countries (e.g., United Kingdom, Netherlands, etc.) are involved in the survey. A few South American countries (e.g., Brazil, Chile, etc.) and African countries (e.g., South Africa, etc.) are also included in the survey based on the locations of respondents' organizations' business operations.
%\begin{figure}
%  \centering
 % \includegraphics[width=0.9\linewidth]{country1.pdf}
 % \caption{The number of participants' organizations in various countries. Since one organization may operate in more than one country, one participant may select more than one country.}
 % \label{country}
%\end{figure}

\section{Our Findings}

Based on the responses from survey participants \footnote{The IDs of participants are used to refer to the survey participant (P). }, this section describes our findings from the survey on the Common Criteria adoption barriers and the potential incentive strategies to encourage the adoption of the Common Criteria.

\begin{figure*}
  \centering
  \includegraphics[width=\linewidth]{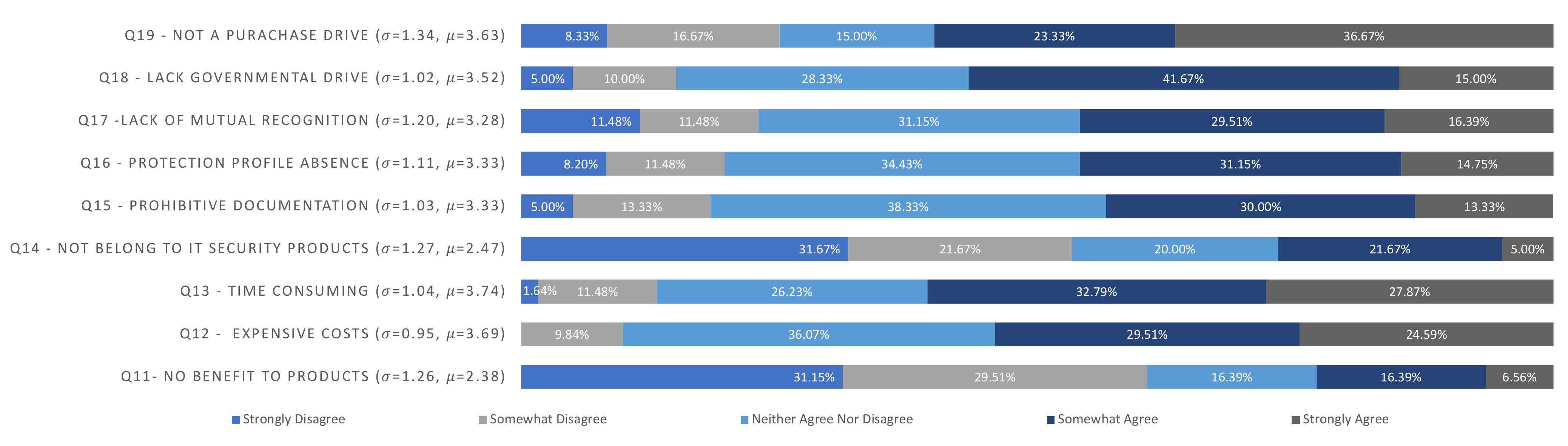}
  \caption{The distribution of agreement level for the identified adoption barriers (i.e., Q11 - Q19). The semantic level of agreement is measured using a five-point scale, where each segment represents the probability of respondents who select that level of agreement (\textit{N}=60).}
  \label{barrier_sta}
\end{figure*}
\subsection{The Common Criteria Adoption Barriers}

We firstly discuss the barriers that discourage organizations from adopting the Common Criteria certification. These adoption barriers are identified from the closed-ended Q11 - Q19 and open-ended Q20. We show the responses to the closed-ended questions on the Common Criteria adoption barriers in Figure \ref{barrier_sta}. Specifically, we identify the following seven adoption barriers that are common to organizations: (1) absence of technology category in the Common Criteria portal; (2) time consuming and not up-to-date; (3) lack of mutual recognition; (4) lack of security evaluation experience; (5) expensive costs; (6) not a key driver for purchase decisions; (7) lack of governmental drive. These adoption barriers will be discussed in detail below.     
%By investigating the Common Criteria adoption barriers, we understand the process of organizations making decisions when they purchase and produce ICT products with security functionalities.
\subsubsection{Absence of Technology Category}

The Common Criteria uses a framework in which the vendors and purchasers can specify their security functional and assurance requirements for the ICT products. The Common Criteria portal \cite{CC_portal} archives the published Protection Profiles and Certified Products under a broad range of categories and diverse technology types. Although the Common Criteria currently cover fifteen categories, there are still 14.75\% participants who strongly agree and 31.15\% participants who somewhat agree the absence of approved Protection Profiles for the category of the products makes it challenging to obtain the Common Criteria Certification, based on the responses for Q16. In particular, participant with the ID number 2 (P2) specified:

\begin{quote}
\textit{``Protection Profiles do not exist for some new technology used by the government such as SD-WAN (i.e., Software-Defined Wide Area Network)"}. 
\end{quote}
In addition, the increasing adoption of emerging technologies motivates the users on bringing in the potential categories of the Common Criteria \cite{DACCA_report1}. The lack of emerging technology category in Common Criteria is an adoption barrier under the reason of absence of technology category. P273 explained:
\begin{quote}
\textit{``There is no Common Criteria method for evaluating products which have some security functionality delivered partially or fully from the cloud."} 
\end{quote}

%In addition, the increasing adoption of emerging technologies motivates the users on introducing the potential categories of the Common Criteria \cite{DACCA_report1}.

 Components required to operate and manage enterprise ICT environments account for a large proportion of the Common Criteria certified products. For example, according to the statistics on Common Criteria certified products \cite{CC_certifiedProductstatistics}, by December 2021, there are 580 certified products under the category of the Integrated Circuits (ICs), smart cards, and smart card-related devices and systems, which takes up the most significant percentage of certified products. However, some organizations found that it is hard to certify their products, compared with the ICT infrastructure products. P22 shared their experiences:
\begin{quote}

\textit{``Organization produces building control systems, including HVAC (i.e., Heating, Ventilation, Air Conditioning), Fire \& Security, airport management systems, and many IoT products. Historically, the government agencies running the Common Crieria are not interested in these market sectors, remaining focused on infrastructure products to the exclusion of others."}

\end{quote}

\subsubsection{Time Consuming and Not Up-to-Date}
    The Common Criteria evaluation process requires a series of stages, including Security Target evaluation, design evaluation, guidance evaluation, life-cycle evaluation, functional testing, and penetration testing. Besides, the evaluation and testing process needs formal documentation following the Common Criteria convention, which takes time to study and compile  \cite{kaluvuri2014quantitative} \cite{murdoch2012certification}. On average, the time required for the Common Criteria certification is six months to one-year \cite{matheu2020survey}. 
    
    More than half of the participants (60.66\%) agree that the Common Criteria evaluation time is too long compared to the product life cycle (Q13). For products that have a short time-to-market, the lengthy evaluation period hinders the adoption of the Common Criteria. To illustrate, P16 shared their experience on Common Criteria certification:

\begin{quote}
 \textit{ ``Another factor hindering Common Criteria adoption is that the approval is only for a specific release, which is years old by the time certification is obtained...There should be a path to quickly have a new version accepted for certifications."}
\end{quote}

 % A majority of the participants (60.66\%) strongly or somewhat disagree the statement in Q11: \textit{Common Criteria certification does not add any benefits to your products}.  However, some organizations think the evaluation process takes time and cost but adds little security benefit to their products. P273 said:

 %\begin{quote}

 %   \textit{``The higher EAL level requires documentation to be written that primarily is to educate the CC evaluator on how the product works and does not result in any change to the security product nor the customer guides. This drives up costs and time with little security benefit."}
    
  %  \end{quote} 
   As technology changes rapidly, the time-consuming process of Common Criteria evaluation and certification impedes the commercialization of security products in the market. As an example, when the Common Criteria certification process concludes, the technology for manufacturing low-cost IoT devices may have become obsolete. Customers may expect the latest features instead of absolute security assurance. As a proof, P125, as one of the largest multi-disciplinary insurance agencies in Australia, said: \textit{``Clients want latest features, these haven't time for Common Criteria evaluation prior to release..."}. P273 from the telecommunications equipment company had similar concerns: \textit{``The software release cadence of cloud products (e.g., monthly) would not align with Common Criteria certification time-frames."}
%``Clients want latest features, these haven't time for Common Criteria evaluation prior to release, but still undergo vulnerability scanning, automated code reviews, pen testing - so are deemed satisfactory."

\subsubsection{Lack of Mutual Recognition}
The Common Criteria is the driving force for the widest available mutual recognition of secure ICT products. As of 2021, there are 17 certificate authorizing schemes under the Common Criteria \cite{CC_authority}. Country-specific implementation of Common Criteria schemes are different in the flow of evaluation and certification of ICT products \cite{fatima2021survey}. The fragmented landscape of Common Criteria schemes generates the disharmonized perspective for security evaluation. In particular, the US keeps the certified products listed for two years before being archived, while the other counties keep them for five years. Nearly half (45.90\%) of our participants agree that there is a lack of mutual recognition on Common Criteria certification among the countries where your products are sold (Q17). P273 emphasized the different lifetimes for different schemes, \textit{``Different schemes enforce different certificate lifetimes, e.g., NIAP - 2 years, Canada - 5 years etc.''}  

The lack of mutual recognition across diverse security standards is one adoption barrier for security certifications, including the Common Criteria \cite{matheu2020survey}. The comment from one of the ICT product consumers P103 remarked \textit{``Certification overload makes anyone choice hard."} Harmonization of the certification results across different standards is expected from both consumers and vendors to improve the transparency of the security certifications. As the vendor, P16 stated:
\begin{quote}

\textit{``No commonality among related certifications. Why can't testing for FIPS 140-3 be accepted as testing for NDcPP? Why can't NDcPP testing be accepted for EAL2? Why can't NDcPP and EAL2 be combined? I would like to see a requirement table which showed relationships between EAL2, NDcPP, FIPS 140-3, EU Cybersecurity law, California Cybersecurity law..."} 

\end{quote}

\subsubsection{Lack of Security Evaluation Experience}
The Common Criteria standard is somewhat complex and not easy to follow to conduct evaluation in terms of its usability and readability. For example, participant P147 commented \textit{``unclear on the process to undertake evaluation''}, P4 stated \textit{``internal resources unavailability"}, and P14 recommended that \textit{``streamline Common Criteria adoption process will be helpful."} 43.33\% of our participants agreed that the documentation requirements for Common Criteria evaluation are prohibitive, so that it is difficult to obtain the Common Criteria certification (Q15). In response, many product vendors engage consultants to prepare specific evaluation material at the pre-evaluation stage.  The current efforts on major review work are underway by international experts through the International Organization of Standardization (ISO) \cite{fatima2021survey}. This should see an improvement of the Common Criteria for wider adoption from the perspective of usability and readability of the documentation.

However, updating and revising the Common Criteria make the certification process inconsistent when users intend to obtain a Common Criteria certification. A few national evaluation schemes are phased out of using Evaluation Assurance Levels (EALs) and only accept products that claim strict conformance with Protection Profiles approved by them. In fact, only Protection Profile evaluations are currently allowed in the United States. P18 mentioned this difficulty when they tried to certify their product under a particular Common Criteria scheme:
\textit{``Lack of consistency and changing rules in the middle of an evaluation makes it very difficult to properly plan and evaluate our products."}

Furthermore, it is not within the scope of the Common Criteria to detail how cryptography is implemented within the TOE. Instead, national standards, such as FIPS 140-2 \cite{no1fips}, specify the specifications for cryptographic modules, and various standards specify the cryptographic algorithms used. In recent years, the Protection Profile authors have included cryptographic requirements for Common Criteria evaluations that would generally be covered by FIPS 140-2 evaluations, expanding the scope of the Common Criteria by using scheme-specific interpretations. 

In addition, a common phenomenon of the security evaluation is the lack of talent and expertise, especially Common Criteria. The outflow of talent with the security evaluation expertise is happening. According to the response from an Australian technology company P152: 

\begin{quote}
\textit{``... New entrants don't see a career path and prefer to work out how to leave as soon as possible... A small core remains in security evaluations."}
\end{quote}

%Most of the [Company Name] employees hired for security evaluations or evaluation support usually find work completing GRC (i.e., Governance,  Risk  and Compliance) documents, Account Security Officer or IT Security Officer.

\subsubsection{Expensive Costs}

In our survey, 24.5\% of participants strongly agree, 29.5\% somewhat agree, and no participant strongly disagrees that the Common Criteria evaluation costs are too expensive compared to the benefits brought into the evaluated products (Q12).
The evaluation costs to obtain the Common Criteria certification is commonly regarded as a barrier to the Common Criteria certification adoption \cite{baldini2016security}\cite{matheu2020survey}, particularly for companies with a limited budget and low-margin products. ICT products with a low-profit margin may not be able to justify and defray the costs associated with Common Criteria certification, given the market's competitive nature. Based on the investigation on certified products listed on the Common Criteria portal \cite{CC3}, the Common Criteria certification is relatively more likely to be adopted by companies with a high-profit margin and capable of sustaining sensitive government networks. P96 shared their thoughts on the cost of Common Criteria evaluation:
\begin{quote}

\textit{``Common Criteria (and its predecessor) was a nice idea but has always been too expensive..."}
\end{quote}

Generally, in line with the Australia scheme, obtaining the Common Criteria certification involves four steps: \textit{pre-evaluation}, \textit{conduct}, \textit{conclusion}, and \textit{assurance continuity}. In the first step, \textit{pre-evaluation} is essential to ensure the success of the Common Criteria evaluation process, prevent delays, conduct initial assessments, develop the Security Target and the evaluation schedule. This includes the writing of functional, high-level, and low-level design specifications. As a second step, the \textit{conduct} phase verifies any claimed security functionality under the Common Criteria, and any other claimed cryptographic functionality under the specific security standard, such as FIPS 140-2 \cite{Evans2002fips}. The evaluation and certification processes are finalized in the \textit{conclusion} phase. Additionally, \textit{the assurance continuity} phase allows for a minimization in the number of evaluations and the options of extending certification to the updated Target of Evaluation version. 

In general, the costs of an evaluation depends on the security assurance level or Protection Profile conformance claims, as well as on the complexity of the Target of Evaluation \cite{Smith2007trend}. A start-up database company P45 commented on the cost of Common Criteria evaluation:
\begin{quote}

\textit{``We've learnt that the certification cost may be above 100k USD, which is too expensive for startups or companies in early developing stages, like ours."}
\end{quote}

The overall evaluation costs are composed of four components: \textit{internal costs}, \textit{external costs}, \textit{lab fees}, and \textit{certification fees}. The \textit{internal costs} are incurred on preparing deliverables and supporting the evaluators. The \textit{external costs} consist of consultancy fees. The \textit{lab fees} are paid to the evaluation labs, and the \textit{certification fees} are paid to the corresponding certification body if applicable. The cost of conducting complex evaluation activities in laboratories is substantial. A recently estimated average cost for a Common Criteria certification lifecycle is US\$250,000 \cite{matheu2020survey} depending on the evaluation assurance level and re-use of past evaluation effort. The cost of an evaluation against a Protection Profile is relatively inexpensive due to the reduced efforts in developing the evaluation documentation. Because of the heavy cost on the Common Criteria lifecycle, it is challenging to evaluate the products against the Common Criteria standard for organizations. 

\subsubsection{Not a Key Driver for Purchase Decisions}

A majority of the participants (60.66\%) strongly or somewhat disagree with the statement in Q11: Common Criteria certification does not add any benefits to your products. However, when considering issues related to the economic viability of the organizations, 36.67\% of our participants strongly agree that Common Criteria certification seems not to be a key driver for purchasing decisions of commercial customers (Q19). For example, P147 said: \textit{``Unclear on the value it would provide to the business."}
%When our participants mentioned  who produce ICT products, \textit{Common Criteria certification seems not a key driver for purchasing decisions of commercial customers} in the target market of some participants' products. 

For start-up companies or low-margin businesses, the budget for gaining security certifications is limited. As a participant from a start-up company, P27 realized that the organization should take countermeasures to cope with cyber attacks. However, the highest priority for spending time and money is on delivering new functionality:

\begin{quote}

\textit{``...As a startup, our focus is on adding additional functionality and must-have features than on optional ones...Our customers are not that security-focused."}
%We use standardized encryption libraries to avoid the simplest security breaches, as a startup our focus is though on adding up additional functionality and must to have features than on optional ones. As said before our customers are not that security focused
\end{quote}

Additionally, some organizations expect to gain economic benefits from Common Criteria certification in addition to security assurance. P27 further put it: \textit{``It is difficult to forecast revenue associated to a certification to justify certification expenses."} P112 pointed out that the decision to pursue Common Criteria certification for their products is partly determined by market demand:
\begin{quote}

\textit{``Certification of a product must contribute to the commercial viability of the product. This should be done by a combination of measures to change the cost/effort barrier to achieving product certification and to improve the unit price/accessible market/demand for the certified product."} 
\end{quote}
\subsubsection{Lack of Governmental Drive}

Within the Q18 respondents, 28\%  strongly or somewhat agree that there is a lack of governmental drive (e.g., security certification requirements) in their procurement policy for the Common Criteria certification. P111 commented: \textit{``People once cared about Common Criteria certification, but the current PSPF (i.e., Protective Security Policy Framework)/ISM (i.e., Information Security Manual) really don't encourage us."} Some organizations found it hard to seek help and support from government agencies, such as the input from P112: 

\begin{quote}
\textit{``Government and commercial do not see product certification as a key part of the risk management around the selection and implementation of effective security controls. Government agencies are very unhelpful in establishing any form of commercial justification to get a product certified and in getting products through any form of the certification process."}
    
\end{quote}

\subsection{Adoption of the Common Criteria: Recommendations}
According to the response received from our survey participants, the above identified barriers hinder the adoption of the Common Criteria. Security standards have always been considered an effective way to provide cyber assurance, although there are some obstacles to widely embracing these security standards. To drive the broad adoption of the Common Criteria and the wider range of security standards, we next discuss the identified Common Criteria incentive strategies and summarize six categories as follows: (1) guidance, resources, and expertise; (2) governmental incentive; (3) mutual recognition; (4) procedure optimization; (5) extension into emerging technologies; (6) consumers' trust.

\subsubsection{Guidance, Resources, and Expertise}
The guidance and resources on how to begin the Common Criteria assessment process, including developing Protection Profile, preparing Security Target, and testing for evaluation, are highly desirable. Although there are documents that introduce the Common Criteria general model \cite{CCpart2}, security functional requirements \cite{CCpart1}, and security assurance requirements \cite{CCpart3} available on the Common Criteria portal \cite{CC_portal}, many participants found it is hard to initiate the procedures due to the lack of usability and readability of these documents as identified in Section 3. For example, P147 emphasized the need for \textit{``better guidance on when a product should be evaluated, and how to begin the process."}

Besides the Common Criteria portal, another important source of information is the International Common Criteria Conference (ICCC) \cite{ICCC}. It is a technical conference where professionals involved in the Common Criteria exchange their experiences in specification, development, evaluation, certification and approval with regard to the ICT security of products and systems. Although the ICCC is held annually for the community of professionals involved in Common Criteria, the experiences shared on the process of Common Criteria assessment are hard to access through public resources. Based on the situation, some companies choose to utilize \textit{``outsourced service"} (P4) that transfers the tasks to professionals with Common Criteria expertise. Moreover, some companies choose to establish their own information bank. P147 shared their approaches on how to accumulate guidance, resources, and expertise on the Common Criteria:
\begin{quote}

\textit{``[Company name] has its own training material and induction process for trainee evaluators. This material includes specific examples of EAL2+ assurance classes and evaluations. A training package that included completed examples of EAL2+ (including ALC\_FLR.2) assurance classes ``which was intended to be shared with TOE developers" would be ideal as it will allow new entrants to view and understand the source material required as inputs to the assurance classes. It would also allow new [Company name] workers to view and create templates to speed up the creation of assurance classes."}
\end{quote}

\subsubsection{Governmental Incentive}
According to the responses from our participants, government supports and incentives motivate the adoption of Common Criteria evaluations and Common Criteria certified products to a certain extent. As the manufacturers of ICT products, some participants adopt the Common Criteria to meet the government's procurement requirements in order to access the markets. For example, P273 shared one of the reasons for adopting  Common Criteria evaluations:
\begin{quote}

\textit{``As an IT equipment manufacturer, [Company name] adopts the Common Criteria as a market access requirement for various government markets globally."} 
\end{quote}

Consumers of Common Criteria certified products can be categorized based on market sectors: public and private. When it comes to the public sector, government supports can boost the adoption of Common Criteria certifications through the requirement of Common Criteria certified products. For example, establishing policy requirements for the procurement process used by government departments and agencies encourages the Common Criteria adoption. The US government requires Common Criteria certified products for specific applications. This policy encourages vendors to participate in Common Criteria evaluations \cite{Smith2007trend}. P103 shared the ideas on the Common Criteria incentive strategies: \textit{``Government mandate or engagement, critical mass in the market."} Similar comments came from P274 and P27: \textit{``Governments mandate for customer and assistance to vendors to get started"} (P274), \textit{``Government regulation"} (P27).

In addition, government incentives can encourage the adoption of Common Criteria certified products in the private sector. The government's support and incentives would be essential in boosting the uptake of Common Criteria certifications since the vendors could minimize legal risks and gain economic benefits from performing the Common Criteria evaluations. For instance, in Japan, tax deductions are available for businesses that use Common Criteria certified products, which increases the purchase of the certified products \cite{yajimaconsideration}.  \textit{``Government funding"} (P106), \textit{``Government grants"} (P100), \textit{``Sponsorship from government"}(P61) and \textit{``Tax incentive"}(P3) were proposed to encourage the adoption of the Common Criteria from our participants for responding Q21 - What kind of incentive would be helpful for your organization for adopting Common Criteria certification.

\subsubsection{Mutual Recognition}

Globally, there are a variety of cybersecurity standards, including international, national, and industry-specific regulations. Comparing the level of security between different standards is difficult. As P44 stated: \textit{``There are many security certification standards."}. Even for the single Common Criteria standard, it is difficult to achieve the objective of comparability due to the technical nature of the document \cite{matheu2020survey}. To achieve the objective, it was proposed to establish a single comprehensive standard to facilitate mutual recognition among various security standards. For example, P189 remarked:
\begin{quote}

\textit{``I think a mass move toward a single comprehensive standard within [Country name] would strongly influence my organization to re-evaluate the need to maintain a standard that essentially duplicates effort and paraphrases similar criteria to other standards."}
\end{quote}

Evaluation in the future can be made more comparable and harmonized by standardizing evaluation activities. For product comparability, rigorous security metrics can be developed that indicate the level of threats, risks, and security provided by each.
\subsubsection{Procedure Optimization}

Through unified processes and formalized steps, evaluation activities are made more manageable. For example, P159 expected \textit{``Lower barriers to entry and a quicker, more transparent process"} to adopt the Common Criteria. Usually, implementation-independent Protection Profiles in the Common Criteria \cite{lee2010protection} define the security requirements for ICT technology that consumers expect. Independent laboratories then evaluate products to decide if the claimed security properties have been achieved \cite{CCevlauationlab}. Standardization of evaluation and testing procedures in the future will make the certification process more transparent. 

In addition, \textit{``the agile and swift"} (P14) process can be considered as the future direction of the optimization procedure for the Common Criteria. Many participants are expecting the \textit{``faster timeframes"} (P130), \textit{``faster pace with faster path for features updates"} (P156) for the Common Criteria certification. In order to respond quickly with rapid iterations and updates on technology, the Common Criteria need to be continuously updated with requirement discovery and solution improvement through the collaboration of vendors, technical specialists, customers, and governments.

Lastly, our participants are concerned about the costs associated with Common Criteria evaluation and certification, as mentioned in Section 4.1.5. There are several proposed ways of reducing the cost proposed, including \textit{``Reducing gap analysis cost, consulting cost, evaluation voucher, etc."} (P10), \textit{``Free evaluations for two products per company."} (P125), and governmental supports discussed in Section 4.2.2.

\subsubsection{Extension into Emerging Technologies}

ICT security-related technologies and evaluations are covered by the Common Criteria, from functionalities to security assurance. Traditional ICT technology and products, such as ICs, database and network devices, are sufficiently covered and evaluated under the Common Criteria standard in the past decades \cite{herrmann2002using}. In light of emerging technologies, the Common Criteria standard needs to address the evaluation of new technologies, such as blockchain, quantum computing, artificial intelligence, and IoT. This is confirmed by P22 who indicated the importance of: \textit{``extension into IoT and commercial sectors."} Additionally, privacy laws should be observed for high assurance products, such as privacy-preserving authentication.

\subsubsection{Consumer Trust}

The adoption of security-sensitive ICT products relies heavily on the trust the users' place in the security features of these products. The trust of users are considered the driving force behind certifications and cybersecurity standards. P41 stated:
\begin{quote}

\textit{``Adopting Common Criteria certification or not is up to the commercial customers in the target market of the products."}
\end{quote}

Assuring the security of ICT products is a joint endeavour between vendors, technical specialists, customers, and governments that never ends. Through education and information available on the Common Criteria portal \cite{CC_portal} and other platforms, sharing information on the core blocks of Common Criteria evaluation and certification will contribute to this cause. The trustworthiness of ICT products and the Common Criteria can be established if consumers are provided with long-term security assurances regarding the products' security features. C150 identified other factors that influence the purchase of security-enhanced ICT products, aside from certification:
\begin{quote}
    
\textit{
``From the perspective of a consumer of security products, Common Criteria (or similar) is not a factor. It is the real world efficacy of a security product combined with the ability to readily implement, maintain and manage that influence the purchasing decision, not a certification."}
\end{quote}

The implementation of security-by-design in product engineering processes can not only significantly shorten the evaluation and certification process, but also ensure that products are designed from the very beginning to be secure \cite{cybersecurityAgencyofSingapore}. The incremental certification of products for additional functionality and features will be more accessible with the integration of certification and evaluation into the product development process \cite{andrea2020towards}\cite{beznosov2004towards}. Furthermore, the certification itself, accompanied by continuous assurances of products' security to consumers, helps consumers build and strengthen trust in the Common Criteria.

\section{Risk Management Directions for Cyber Assurance}

By investigating the Common Criteria adoption barriers, we understand the process of organizations making decisions when they purchase, produce, and use ICT products with security functionalities. In addition to adopting security standards such as the Common Criteria as discussed in Section 4, there are other risk management approaches to achieve cybersecurity assurance that organizations can take to protect their assets and the data of their employees, business partners, and customers. %An organization can choose to take a proactive approach to cybersecurity, preventing threats before they arise, or a reactive approach, addressing cybersecurity breaches after they have occurred.
We further discuss our investigation of risk management approaches for cyber assurance adopted by the participating organizations in our survey with the proposed future directions for risk management of organizations. 

\subsection{Reactive vs Proactive cybersecurity}
An organization can choose to take a proactive approach to cybersecurity, preventing threats before they arise, or a reactive approach, addressing cybersecurity breaches after they have occurred. Reactive cybersecurity investigates the signs that indicate a data breach has occurred and a cybersecurity incident has been committed \cite{xu2020cybersecurity}. The reactive approach involves responding to cybersecurity incidents in case of further damage \cite{benzel2020cybersecurity}. Proactive cybersecurity investigates the indicators of compromise, which is a broad and overall approach that involves not only specific methods and practices but also a mindset of protecting cybersecurity before the incidents happen in advance \cite{sun2018data}. Below, we categorize the cybersecurity risk management approaches shared by our participants and analyze these approaches from the reactive and proactive points of view respectively.

\subsubsection{Patch Management}
Patching falls under the Essential Eight in the Strategies to Mitigate Cyber Security Incidents of the Australian Cyber Security Centre \cite{essential8}. Patch management is applied to computer systems, applications, cloud infrastructure, and other critical infrastructure
(e.g., industrial control systems) to mitigate cybersecurity incidents.  Once a vendor releases a patch, the patch should be applied in a timeframe commensurate with an organization’s exposure to the security vulnerability and the level of cyber threat the organization is aiming to protect itself against. Once a newly discovered security vulnerability in an internet-facing service is made public, adversaries will likely develop malicious code within 48 hours \cite{patchACA}.

Some of our participants handled patching management in their own way. For example, P229 mentioned that they conducted \textit{``regular review of available updates and patching"} as part of the risk management process, and P134 reviewed \textit{``patch management forums"} to check the feedback on the products prior procurement. A reactive patch is applied in response to an issue that currently affects a system and that needs immediate relief \cite{oraclepatch}. When such a situation occurs, users typically install the most recent patch or patches, which may appear to be capable of resolving the issue. However, in many cases, problems that can occur have already been identified, and patches have already been released. Compared to a reactive patch management strategy, a proactive patch management strategy implies more changes and regularly scheduled maintenance windows to reduce unplanned issues \cite{nicastro2019security}.

\subsubsection{Cyber Risk Profile}

The cyber risk profile is a quantitative approach for assessing cybersecurity risks for an organization, asset, project or individual \cite{sokri2019cyber}. In the absence of cybersecurity certifications, some participants establish a risk profile for the product to manage risks. For example, P193 shared their experience on how to manage risks when there are no cybersecurity certifications for the products:

\begin{quote}
\textit{``...We establish an overall risk profile for the product to consider what information will be stored, processed or communicated using the product, as well as establishing mitigating or alternative controls to manage the absence of the certification."}   
\end{quote}

As a reactive way, the audit data can be retrieved to avoid further damage if cybersecurity incidents have happened \cite{liu2019insider}. Furthermore, the risks are monitored based on the established risk profile. For example, P140 mentioned that the risks are monitored in \textit{``the lifecycle of the supportability of the product"}. The procedures and countermeasures are formulated based on the risk profile to understand the risks, assess them, and mitigate them. Therefore, the organization proactively manages the risks on products, systems, assets, and projects to reduce the likelihood of cyber attacks, as illustrated by P156's example:

\begin{quote}
\textit{``Evaluate the risk profile versus benefit for the product in question and put in place commensurate controls and standard operating procedures to minimize the security risk."}
\end{quote}

It is worth mentioning that through the secure sharing of risk profiles across organizations, the information on risk management can be shared to improve cyber resilience. For example, the Open Science Cyber Risk
Profile (OSCRP) aims to help improve IT security for open science
projects \cite{peisert2017open} for scientists and IT professionals, which serves to bridge the communication gap between scientists and IT security professionals and allows for the effective management of risks to open science caused by IT security threats.  
\subsubsection{Self and In-house Evaluations}

Independent testing, such as \textit{``penetration testing''} (P11), \textit{``own vulnerability scanning''} (P125), \textit{``in-house examination''} (P104), was adopted by our survey participants for cyber assurance based on the survey. Through independent testing, ICT products are tested with white-hat hackers to find exploitable vulnerabilities. In addition, potential exploits can be closed with the help of penetration testing in the proactive approach. The mitigation recommendations and strategies will be decided to conduct risk management within organizations based on the testing results. P273 shared their experience on their self and in-house evaluations:
\begin{quote}
\textit{``For cloud products, we conduct a detailed cloud security assessment based on the business criticality of information classification of its use within [Company name]. For on-prem products, we conduct in-house evaluations prior to deployment.''}
\end{quote}

\subsubsection{Cybersecurity Insurance}

Cyber risk management is imperative due to the significant economic impacts and increased media attention \cite{biener2015insurability}. In the light of the need for improving risk management within organizations, cybersecurity insurance companies have been developing steadily in recent years \cite{talesh2018data}. For instance, BizCover \cite{BizCover} is an insurance company that will cover expenses on cybersecurity
incidents. In the context of cybersecurity insurance, the cyber risk is tagged with a price, which creates incentive for risk-appropriate behavior. In addition, by simply applying for cybersecurity insurance, the organizations become more aware of self-protective awareness against the cyber threats. Cybersecurity insurance assists users in taking a proactive approach to cybersecurity in addition to potentially covering the financial cost of dealing with cybersecurity attacks in the reactive way \cite{evans2016human}.

Some of our participants, especially small businesses, chose to use insurance companies to seek assurances in the cybersecurity aspect. For example, P7 is a global IT company with the size between 51-250 employees mentioned that they employed \textit{``insurance companies"} to manage risks associated with potentially poor implementation of security functionality within the products in the absence of IT products with a security certification standard and used \textit{``insurance companies"} to seek cyber assurance. 

However, there are a number of difficulties that restrict the development of cybersecurity insurance, including loss occurrence, information asymmetries, and the limits of insurance coverage \cite{biener2015insurability}. With the increasing market development, the risk information pool for cybersecurity insurance will become more extensive with more available data. Therefore, sharing data through national regulators or international associations will improve insurance risk assessments and insurance market efficiency.

 \subsubsection{Cybersecurity Awareness Education}
According to MediaPRO's annual privacy and security awareness report, 85\% of finance workers lacked knowledge around data privacy and cybersecurity \cite{awarenessreport}. Besides the professional education on evaluation as introduced in Section 4.1.4, cybersecurity uncertainty will be mitigated by keeping employees up to date on the latest threat intelligence and attack methods. In addition to reducing stress, security training helps eliminate risky behaviours and establishes a culture of cybersecurity in the workplace regarding security standards.  

Based on the responses from our questionnaires, many organizations provided cybersecurity training to their staff to establish their awareness. For example, P67 said that they offered \textit{``policies/procedures and staff cybersecurity training"}, and P229 shared that they trained their staff with \textit{``industry best practices guides"} and \textit{``previous implementation experience."}

Besides the training on cybersecurity standards, the best practices, and standards, making sure users know how to spot the tell-tale signs and tricks of fraudsters will enable them to avoid social engineering and other phishing attacks.

%\subsection{Other cyber Assurance Approaches}

%\textbf{Vendors collaboration. }

%\textbf{Government recommendations.}

%\textbf{Best practices and certifications. }

%\textbf{Brand reputation.}

%\textbf{Contract}

%\textbf{Reference check}

%\textbf{Cyber insurance}

%\textbf{Self-trust}

%\textbf{No seeking assurance}

\subsection{Multi-layered Risk  Management}

Although a few participants indicated that \textit{``never attempted to"} (P3) seek cybersecurity assurances with the product and \textit{``not used"} (P246) risk management. In most cases, our participants took steps to ensure cybersecurity and prevent potential threats. Rather than relying on a single approach to managing risks, most of our participants adopted multiple risk management strategies. For example, P262 introduced how to obtain cyber assurance within the organization when accessing the ICT products:

\begin{quote}

\textit{``For products and services that we have access to, we will use a mix of audit and technical testing capabilities to obtain an appropriate level of assurance. The degree of this activity is dependant on the risk."}

\end{quote}

In order to effectively manage risk, organizations need a systemic multi-layered approach that crosses multiple business units, departments and processes, touching every individual, machine and element within the organization. Below, we discuss how to obtain cyber assurance through the multi-layered risk management based on the responses from our participants.

Firstly, in the process of product design and implementation, practical risk management tools and approaches are employed through assurance activities. For example, P78 mentioned that they utilized \textit{``vulnerability management and mitigation controls where possible"} to manage risks within the products during the product implementation process.  Organizations also follow \textit{``best practices recommended by governments"} (P267) and \textit{``industry best practices guide"} (P229) in this process, such as the implementation of security-by-design as demonstrated in Section 4.2.6. P119 shared their activities in this process: 
\begin{quote}

\textit{``Through detailed threat modeling and analysis and designing required mitigations. Standard cybersecurity risk management practices are then used as a mechanism to provide required assurance."} 

\end{quote}

%In the age of technology, ICT products, tools and resources are used everyday. Therefore s
Secondly, organizations need to protect their cybersecurity and improve their cyber resilience when selecting and utilizing these ICT technologies. Usually, organizations combine a number of available data to choose ICT products. Many participants preferred to choose products with \textit{``trusted brands"} (P143) and \textit{``reputable companies"} (P75). Besides reference checks on the products, security certification is an essential element to check for the purchasers. For example, P83 declared: \textit{``We need to see the certification from the vendors".} In addition, some participants conducted \textit{``own independent testing"} (P159) or \textit{``3rd party efficacy test"} (P150) before procurement and monitor the products during use. P57 shared their activities before procurement to achieve cybersecurity assurance:  

\begin{quote}
    \textit{``Generally achieved through supply chain due diligence, selecting preferred and trusted suppliers and by conducting internal assessment and suitability validation of the products as fit for purpose and being with the risk thresholds of the organization."}
\end{quote}

Thirdly, cybersecurity attacks and incidents are inevitable nowadays. Organizations prepare for the worst before incidents occur and try their best to reduce losses if incidents have already occurred. Another layer of risk management relates to incident management. Audit data should be traceable to help organizations reflect on the attacks to avoid similar attacks in the future. Some of our participants chose to utilize an insurance company as shown in 5.1.4. In addition, setting up a response team to prepare for mitigation recommendations and actions is necessary to bounce back after cybersecurity incidents. 

Furthermore, cybersecurity issues are not limited to the IT department. They pose a significant threat to business continuity and reputation and threaten every aspect of an organization. Security awareness within the company helps employees understand cyber hygiene that refers to the practices for ensuring the safe handling of data and for securing networks \cite{cain2018exploratory}. Educating employees about the security risks associated with their actions via email and the web reduces the chances of being attacked \cite{zhang2021systematic}. As demonstrated in 5.1.5, some of the staff from the organizations of our participants were provided with cybersecurity training. Furthermore, since cybersecurity is a cross-functional concern, the organizations sometimes need to work with external entities to share information on cybersecurity to improve cyber resilience. Hence, besides cybersecurity awareness education, the C-suite level plays an imperative role in establishing a cybersecured organization \cite{sanders2016embedding}.

\section{Conclusion}

In this work, we presented the results of our survey on the Common Criteria adoption and approaches to ensuring cyber assurance for organizations. To determine if organizations have concerns related to cybersecurity regulatory issues as well as to determine organizations' attitudes towards being measured against cybersecurity standards, seven adoption barriers of security standards and certifications are identified. The results of our study inform our recommendations for promoting Common Criteria adoption and broader cybersecurity standards and certifications. Aside from the use of cybersecurity standards and certifications to select secure ICT products, we investigate how organizations pursue cyber assurance and their adopted strategies. We hope the findings and recommendations we have made help researchers, organizations, and regulators raise concerns among academia and industry about the importance of cybersecurity standards and certifications. Beyond cybersecurity standards and certifications, the survey presents insights and directions on risk management, in the hope of inspiring organizations to achieve cyber assurance.

\section*{Acknowledgements}
The work has been supported by the Cyber Security Research Centre Limited whose activities are partially funded by the Australian Government’s Cooperative Research Centres Programme.

{\footnotesize \bibliographystyle{acm}
\bibliography{sample}}

\end{document}